\documentclass[lettersize,journal]{IEEEtran}
\usepackage{cite}
\usepackage{amsmath,amssymb,amsfonts}
\usepackage{algorithmic}
\usepackage{graphicx}
\usepackage{textcomp}
\usepackage{xcolor}
\usepackage{caption}
\usepackage{subcaption}
\usepackage{multicol}
\usepackage{multirow}
\usepackage[normalem]{ulem}
\usepackage{lipsum} 
\usepackage{soul}
\usepackage{epsfig}
\usepackage{epstopdf}
\usepackage{graphics}
\usepackage{balance}
\usepackage{colortbl}

\usepackage[acronyms,nonumberlist,nopostdot,nomain,nogroupskip,acronymlists={hidden}]{glossaries}
\newglossary[algh]{hidden}{acrh}{acnh}{Hidden Acronyms}
\newacronym{3gpp}{3GPP}{3rd Generation Partnership Project}
\newacronym{4g}{4G}{4th generation}
\newacronym{5g}{5G}{5th generation}
\newacronym{5gc}{5GC}{5G Core}
\newacronym{adc}{ADC}{Analog to Digital Converter}
\newacronym{ae}{AE}{Autoencoder}
\newacronym{aerpaw}{AERPAW}{Aerial Experimentation and Research Platform for Advanced Wireless}
\newacronym{ai}{AI}{Artificial Intelligence}
\newacronym{aimd}{AIMD}{Additive Increase Multiplicative Decrease}
\newacronym{am}{AM}{Acknowledged Mode}
\newacronym{amc}{AMC}{Adaptive Modulation and Coding}
\newacronym{amf}{AMF}{Access and Mobility Management Function}
\newacronym{aml}{AML}{Adversarial Machine Learning}
\newacronym{aops}{AOPS}{Adaptive Order Prediction Scheduling}
\newacronym{api}{API}{Application Programming Interface}
\newacronym{apn}{APN}{Access Point Name}
\newacronym{ap}{AP}{application protocol}
\newacronym{aqm}{AQM}{Active Queue Management}
\newacronym{ausf}{AUSF}{Authentication Server Function}
\newacronym{avc}{AVC}{Advanced Video Coding}
\newacronym{awgn}{AGWN}{Additive White Gaussian Noise}
\newacronym{balia}{BALIA}{Balanced Link Adaptation Algorithm}
\newacronym{bbu}{BBU}{Base Band Unit}
\newacronym{bdp}{BDP}{Bandwidth-Delay Product}
\newacronym{ber}{BER}{Bit Error Rate}
\newacronym{bf}{BF}{Beamforming}
\newacronym{bler}{BLER}{Block Error Rate}
\newacronym{brr}{BRR}{Bayesian Ridge Regressor}
\newacronym{bs}{BS}{Base Station}
\newacronym{bsr}{BSR}{Buffer Status Report}
\newacronym{bss}{BSS}{Business Support System}
\newacronym{ca}{CA}{Carrier Aggregation}
\newacronym{caas}{CaaS}{Connectivity-as-a-Service}
\newacronym{cb}{CB}{Code Block}
\newacronym{cc}{CC}{Congestion Control}
\newacronym{ccid}{CCID}{Congestion Control ID}
\newacronym{cco}{CC}{Carrier Component}
\newacronym{cdd}{CDD}{Cyclic Delay Diversity}
\newacronym{cdf}{CDF}{Cumulative Distribution Function}
\newacronym{cdn}{CDN}{Content Distribution Network}
\newacronym{cn}{CN}{Core Network}
\newacronym{codel}{CoDel}{Controlled Delay Management}
\newacronym{comac}{COMAC}{Converged Multi-Access and Core}
\newacronym{cord}{CORD}{Central Office Re-architected as a Datacenter}
\newacronym{cornet}{CORNET}{COgnitive Radio NETwork}
\newacronym{cosmos}{COSMOS}{Cloud Enhanced Open Software Defined Mobile Wireless Testbed for City-Scale Deployment}
\newacronym{cots}{COTS}{Commercial Off-the-Shelf}
\newacronym{cp}{CP}{Control Plane}
\newacronym{cpu}{CPU}{Central Processing Unit}
\newacronym{cqi}{CQI}{Channel Quality Information}
\newacronym{cr}{CR}{Cognitive Radio}
\newacronym{cran}{CRAN}{Cloud \gls{ran}}
\newacronym{crs}{CRS}{Cell Reference Signal}
\newacronym{csi}{CSI}{Channel State Information}
\newacronym{csirs}{CSI-RS}{Channel State Information - Reference Signal}
\newacronym{cu}{CU}{Central Unit}
\newacronym{d2tcp}{D$^2$TCP}{Deadline-aware Data center TCP}
\newacronym{d3}{D$^3$}{Deadline-Driven Delivery}
\newacronym{dac}{DAC}{Digital to Analog Converter}
\newacronym{dag}{DAG}{Directed Acyclic Graph}
\newacronym{das}{DAS}{Distributed Antenna System}
\newacronym{dash}{DASH}{Dynamic Adaptive Streaming over HTTP}
\newacronym{dc}{DC}{Dual Connectivity}
\newacronym{dccp}{DCCP}{Datagram Congestion Control Protocol}
\newacronym{dce}{DCE}{Direct Code Execution}
\newacronym{dci}{DCI}{Downlink Control Information}
\newacronym{dctcp}{DCTCP}{Data Center TCP}
\newacronym{dl}{DL}{Downlink}
\newacronym{dmr}{DMR}{Deadline Miss Ratio}
\newacronym{dmrs}{DMRS}{DeModulation Reference Signal}
\newacronym{dnn}{DNN}{Deep Neural Network}
\newacronym{drlcc}{DRL-CC}{Deep Reinforcement Learning Congestion Control}
\newacronym{drs}{DRS}{Discovery Reference Signal}
\newacronym{du}{DU}{Distributed Unit}
\newacronym{e2e}{E2E}{end-to-end}
\newacronym{ecaas}{ECaaS}{Edge-Cloud-as-a-Service}
\newacronym{ecn}{ECN}{Explicit Congestion Notification}
\newacronym{edf}{EDF}{Earliest Deadline First}
\newacronym{embb}{eMBB}{Enhanced Mobile Broadband}
\newacronym{empower}{EMPOWER}{EMpowering transatlantic PlatfOrms for advanced WirEless Research}
\newacronym{enb}{eNB}{evolved Node Base}
\newacronym{endc}{EN-DC}{E-UTRAN-\gls{nr} \gls{dc}}
\newacronym{epc}{EPC}{Evolved Packet Core}
\newacronym{eps}{EPS}{Evolved Packet System}
\newacronym{es}{ES}{Edge Server}
\newacronym{etsi}{ETSI}{European Telecommunications Standards Institute}
\newacronym[firstplural=Estimated Times of Arrival (ETAs)]{eta}{ETA}{Estimated Time of Arrival}
\newacronym{eutran}{E-UTRAN}{Evolved Universal Terrestrial Access Network}
\newacronym{faas}{FaaS}{Function-as-a-Service}
\newacronym{fapi}{FAPI}{Functional Application Platform Interface}
\newacronym{fdd}{FDD}{Frequency Division Duplexing}
\newacronym{fdm}{FDM}{Frequency Division Multiplexing}
\newacronym{fdma}{FDMA}{Frequency Division Multiple Access}
\newacronym{fed4fire}{FED4FIRE+}{Federation 4 Future Internet Research and Experimentation Plus}
\newacronym{fir}{FIR}{finite impulse response}
\newacronym{fit}{FIT}{Future \acrlong{iot}}
\newacronym{fpga}{FPGA}{Field Programmable Gate Array}
\newacronym{fr2}{FR2}{Frequency Range 2}
\newacronym{fs}{FS}{Fast Switching}
\newacronym{fscc}{FSCC}{Flow Sharing Congestion Control}
\newacronym{ftp}{FTP}{File Transfer Protocol}
\newacronym{fw}{FW}{Flow Window}
\newacronym{ge}{GE}{Gaussian Elimination}
\newacronym{gnb}{gNB}{Next Generation Node Base}
\newacronym{gop}{GOP}{Group of Pictures}
\newacronym{gpr}{GPR}{Gaussian Process Regressor}
\newacronym{gpu}{GPU}{Graphics Processing Unit}
\newacronym{gtp}{GTP}{GPRS Tunneling Protocol}
\newacronym{gtpc}{GTP-C}{GPRS Tunnelling Protocol Control Plane}
\newacronym{gtpu}{GTP-U}{GPRS Tunnelling Protocol User Plane}
\newacronym{gtpv2c}{GTPv2-C}{\gls{gtp} v2 - Control}
\newacronym{gw}{GW}{Gateway}
\newacronym{harq}{HARQ}{Hybrid Automatic Repeat reQuest}
\newacronym{hetnet}{HetNet}{Heterogeneous Network}
\newacronym{hh}{HH}{Hard Handover}
\newacronym{hol}{HOL}{Head-of-Line}
\newacronym{hqf}{HQF}{Highest-quality-first}
\newacronym{hss}{HSS}{Home Subscription Server}
\newacronym{http}{HTTP}{HyperText Transfer Protocol}
\newacronym{ia}{IA}{Initial Access}
\newacronym{iab}{IAB}{Integrated Access and Backhaul}
\newacronym{ic}{IC}{Incident Command}
\newacronym{ietf}{IETF}{Internet Engineering Task Force}
\newacronym{imsi}{IMSI}{International Mobile Subscriber Identity}
\newacronym{imt}{IMT}{International Mobile Telecommunication}
\newacronym{idn}{IDN}{Inference Delivery Networks}
\newacronym{iot}{IoT}{Internet of Things}
\newacronym{ip}{IP}{Internet Protocol}
\newacronym{itu}{ITU}{International Telecommunication Union}
\newacronym{kpi}{KPI}{Key Performance Indicator}
\newacronym{kpm}{KPM}{Key Performance Measurement}
\newacronym{kvm}{KVM}{Kernel-based Virtual Machine}
\newacronym{los}{LOS}{Line-of-Sight}
\newacronym{lsm}{LSM}{Link-to-System Mapping}
\newacronym{lstm}{LSTM}{Long Short Term Memory}
\newacronym{lte}{LTE}{Long Term Evolution}
\newacronym{lxc}{LXC}{Linux Container}
\newacronym{m2m}{M2M}{Machine to Machine}
\newacronym{mac}{MAC}{Medium Access Control}
\newacronym{manet}{MANET}{Mobile Ad Hoc Network}
\newacronym{mano}{MANO}{management~and orchestration}
\newacronym{mc}{MC}{Multi-Connectivity}
\newacronym{mcc}{MCC}{Mobile Cloud Computing}
\newacronym{mchem}{MCHEM}{Massive Channel Emulator}
\newacronym{mcs}{MCS}{Modulation and Coding Scheme}
\newacronym{mec}{MEC}{Multi-access Edge Computing}
\newacronym{mec2}{MEC}{Mobile Edge Cloud}
\newacronym{mfc}{MFC}{Mobile Fog Computing}
\newacronym{mi}{MI}{Mutual Information}
\newacronym{mib}{MIB}{Master Information Block}
\newacronym{miesm}{MIESM}{Mutual Information Based Effective SINR}
\newacronym{mimo}{MIMO}{Multiple Input, Multiple Output}
\newacronym{ml}{ML}{Machine Learning}
\newacronym{mlr}{MLR}{Maximum-local-rate}
\newacronym[plural=\gls{mme}s,firstplural=Mobility Management Entities (MMEs)]{mme}{MME}{Mobility Management Entity}
\newacronym{mmtc}{mMTC}{Massive Machine-Type Communications}
\newacronym{mmwave}{mmWave}{millimeter wave}
\newacronym{mpdccp}{MP-DCCP}{Multipath Datagram Congestion Control Protocol}
\newacronym{mptcp}{MPTCP}{Multipath TCP}
\newacronym{mr}{MR}{Maximum Rate}
\newacronym{mrdc}{MR-DC}{Multi \gls{rat} \gls{dc}}
\newacronym{mse}{MSE}{Mean Square Error}
\newacronym{mss}{MSS}{Maximum Segment Size}
\newacronym{mt}{MT}{Mobile Termination}
\newacronym{mtd}{MTD}{Machine-Type Device}
\newacronym{mtu}{MTU}{Maximum Transmission Unit}
\newacronym{mumimo}{MU-MIMO}{Multi-user \gls{mimo}}
\newacronym{mvno}{MVNO}{Mobile Virtual Network Operator}
\newacronym{nalu}{NALU}{Network Abstraction Layer Unit}
\newacronym{nas}{NAS}{Non-Access Stratum}
\newacronym{nbiot}{NB-IoT}{Narrow Band IoT}
\newacronym{nfv}{NFV}{Network Function Virtualization}
\newacronym{nfvi}{NFVI}{Network Function Virtualization Infrastructure}
\newacronym{nic}{NIC}{Network Interface Card}
\newacronym{nlos}{NLOS}{Non-Line-of-Sight}
\newacronym{no}{NO}{Network Operator}
\newacronym{now}{NOW}{Non Overlapping Window}
\newacronym{nsm}{NSM}{Network Service Mesh}
\newacronym[type=hidden]{nr}{NR}{New Radio}
\newacronym{nrf}{NRF}{Network Repository Function}
\newacronym{nsa}{NSA}{Non Stand Alone}
\newacronym{nse}{NSE}{Network Slicing Engine}
\newacronym{nssf}{NSSF}{Network Slice Selection Function}
\newacronym{o2i}{O2I}{Outdoor to Indoor}
\newacronym{oai}{OAI}{OpenAirInterface}
\newacronym{oaicn}{OAI-CN}{\gls{oai} \acrlong{cn}}
\newacronym{oairan}{OAI-RAN}{\acrlong{oai} \acrlong{ran}}
\newacronym{oam}{OAM}{Operations, Administration and Maintenance}
\newacronym{ofdm}{OFDM}{Orthogonal Frequency Division Multiplexing}
\newacronym{olia}{OLIA}{Opportunistic Linked Increase Algorithm}
\newacronym{omec}{OMEC}{Open Mobile Evolved Core}
\newacronym{onap}{ONAP}{Open Network Automation Platform}
\newacronym{onf}{ONF}{Open Networking Foundation}
\newacronym{onos}{ONOS}{Open Networking Operating System}
\newacronym{oom}{OOM}{\gls{onap} Operations Manager}
\newacronym{opnfv}{OPNFV}{Open Platform for \gls{nfv}}
\newacronym[type=hidden]{oran}{O-RAN}{Open \gls{ran}}
\newacronym{orbit}{ORBIT}{Open-Access Research Testbed for Next-Generation Wireless Networks}
\newacronym{os}{OS}{Operating System}
\newacronym{oss}{OSS}{Operations Support System}
\newacronym{pa}{PA}{Position-aware}
\newacronym{pase}{PASE}{Prioritization, Arbitration, and Self-adjusting Endpoints}
\newacronym{pawr}{PAWR}{Platforms for Advanced Wireless Research}
\newacronym{pbch}{PBCH}{Physical Broadcast Channel}
\newacronym{pcef}{PCEF}{Policy and Charging Enforcement Function}
\newacronym{pcfich}{PCFICH}{Physical Control Format Indicator Channel}
\newacronym{pcrf}{PCRF}{Policy and Charging Rules Function}
\newacronym{pdcch}{PDCCH}{Physical Downlink Control Channel}
\newacronym{pdcp}{PDCP}{Packet Data Convergence Protocol}
\newacronym{pdsch}{PDSCH}{Physical Downlink Shared Channel}
\newacronym{pdu}{PDU}{Packet Data Unit}
\newacronym{pf}{PF}{Proportional Fair}
\newacronym{pgw}{PGW}{Packet Gateway}
\newacronym{phich}{PHICH}{Physical Hybrid ARQ Indicator Channel}
\newacronym{phy}{PHY}{Physical}
\newacronym{pmch}{PMCH}{Physical Multicast Channel}
\newacronym{pmi}{PMI}{Precoding Matrix Indicators}
\newacronym{powder}{POWDER}{Platform for Open Wireless Data-driven Experimental Research}
\newacronym{ppo}{PPO}{Proximal Policy Optimization}
\newacronym{ppp}{PPP}{Poisson Point Process}
\newacronym{prach}{PRACH}{Physical Random Access Channel}
\newacronym{prb}{PRB}{Physical Resource Block}
\newacronym{psnr}{PSNR}{Peak Signal to Noise Ratio}
\newacronym{pss}{PSS}{Primary Synchronization Signal}
\newacronym{pucch}{PUCCH}{Physical Uplink Control Channel}
\newacronym{pusch}{PUSCH}{Physical Uplink Shared Channel}
\newacronym{qam}{QAM}{Quadrature Amplitude Modulation}
\newacronym{qci}{QCI}{\gls{qos} Class Identifier}
\newacronym{qoe}{QoE}{Quality of Experience}
\newacronym{qos}{QoS}{Quality of Service}
\newacronym{quic}{QUIC}{Quick UDP Internet Connections}
\newacronym{rach}{RACH}{Random Access Channel}
\newacronym{ran}{RAN}{Radio Access Network}
\newacronym[firstplural=Radio Access Technologies (RATs)]{rat}{RAT}{Radio Access Technology}
\newacronym{rcn}{RCN}{Research Coordination Network}
\newacronym{rec}{REC}{Radio Edge Cloud}
\newacronym{red}{RED}{Random Early Detection}
\newacronym{renew}{RENEW}{Reconfigurable Eco-system for Next-generation End-to-end Wireless}
\newacronym{rf}{RF}{Radio Frequency}
\newacronym{rfc}{RFC}{Request for Comments}
\newacronym{rfr}{RFR}{Random Forest Regressor}
\newacronym{ric}{RIC}{\gls{ran} Intelligent Controller}
\newacronym{rlc}{RLC}{Radio Link Control}
\newacronym{rlf}{RLF}{Radio Link Failure}
\newacronym{rlnc}{RLNC}{Random Linear Network Coding}
\newacronym{rmr}{RMR}{RIC Message Router}
\newacronym{rmse}{RMSE}{Root Mean Squared Error}
\newacronym{rnis}{RNIS}{Radio Network Information Service}
\newacronym{rr}{RR}{Round Robin}
\newacronym{rrc}{RRC}{Radio Resource Control}
\newacronym{rrm}{RRM}{Radio Resource Management}
\newacronym{rru}{RRU}{Remote Radio Unit}
\newacronym{rs}{RS}{Remote Server}
\newacronym{rsrp}{RSRP}{Reference Signal Received Power}
\newacronym{rsrq}{RSRQ}{Reference Signal Received Quality}
\newacronym{rss}{RSS}{Received Signal Strength}
\newacronym{rssi}{RSSI}{Received Signal Strength Indicator}
\newacronym{rt}{RT}{Real-time}
\newacronym{rtt}{RTT}{Round Trip Time}
\newacronym{ru}{RU}{Radio Unit}
\newacronym{rw}{RW}{Receive Window}
\newacronym{rx}{RX}{Receiver}
\newacronym{s1ap}{S1AP}{S1 Application Protocol}
\newacronym{sa}{SA}{Security Association}
\newacronym{sack}{SACK}{Selective Acknowledgment}
\newacronym{sap}{SAP}{Service Access Point}
\newacronym{sc2}{SC2}{Spectrum Collaboration Challenge}
\newacronym{scef}{SCEF}{Service Capability Exposure Function}
\newacronym{sch}{SCH}{Secondary Cell Handover}
\newacronym{scoot}{SCOOT}{Split Cycle Offset Optimization Technique}
\newacronym{sctp}{SCTP}{Stream Control Transmission Protocol}
\newacronym{sdap}{SDAP}{Service Data Adaptation Protocol}
\newacronym{sdk}{SDK}{Software Development Kit}
\newacronym{sdm}{SDM}{Space Division Multiplexing}
\newacronym{sdma}{SDMA}{Spatial Division Multiple Access}
\newacronym{sdn}{SDN}{Software-defined Networking}
\newacronym{sdr}{SDR}{Software-defined Radio}
\newacronym{seba}{SEBA}{SDN-Enabled Broadband Access}
\newacronym{sgsn}{SGSN}{Serving GPRS Support Node}
\newacronym{sgw}{SGW}{Service Gateway}
\newacronym{si}{SI}{Study Item}
\newacronym{sib}{SIB}{Secondary Information Block}
\newacronym{sinr}{SINR}{Signal to Interference plus Noise Ratio}
\newacronym{sip}{SIP}{Session Initiation Protocol}
\newacronym{siso}{SISO}{Single Input, Single Output}
\newacronym{sla}{SLA}{service level agreement}
\newacronym{sm}{SM}{Service Model}
\newacronym{smf}{SMF}{Session Management Function}
\newacronym{smo}{SMO}{Service Management and Orchestration}
\newacronym{sms}{SMS}{Short Message Service}
\newacronym{smsgmsc}{SMS-GMSC}{\gls{sms}-Gateway}
\newacronym{snr}{SNR}{Signal-to-Noise-Ratio}
\newacronym{son}{SON}{Self-Organizing Network}
\newacronym{sptcp}{SPTCP}{Single Path TCP}
\newacronym{srb}{SRB}{Service Radio Bearer}
\newacronym{srn}{SRN}{Standard Radio Node}
\newacronym{srs}{SRS}{Sounding Reference Signal}
\newacronym{ss}{SS}{Synchronization Signal}
\newacronym{sss}{SSS}{Secondary Synchronization Signal}
\newacronym{st}{ST}{Spanning Tree}
\newacronym{svc}{SVC}{Scalable Video Coding}
\newacronym{tb}{TB}{Transport Block}
\newacronym{tcp}{TCP}{Transmission Control Protocol}
\newacronym{tdd}{TDD}{Time Division Duplexing}
\newacronym{tdm}{TDM}{Time Division Multiplexing}
\newacronym{tdma}{TDMA}{Time Division Multiple Access}
\newacronym{tfl}{TfL}{Transport for London}
\newacronym{tfrc}{TFRC}{TCP-Friendly Rate Control}
\newacronym{tft}{TFT}{Traffic Flow Template}
\newacronym{tgen}{TGEN}{Traffic Generator}
\newacronym{tip}{TIP}{Telecom Infra Project}
\newacronym{tm}{TM}{Transparent Mode}
\newacronym[plural=Telcos,firstplural=Telecommunications Companies (Telcos)]{to}{Telco}{Telecommunications Company}
\newacronym{tr}{TR}{Technical Report}
\newacronym{trp}{TRP}{Transmitter Receiver Pair}
\newacronym{ts}{TS}{Technical Specification}
\newacronym{tti}{TTI}{Transmission Time Interval}
\newacronym{ttt}{TTT}{Time-to-Trigger}
\newacronym{tx}{TX}{Transmitter}
\newacronym{uas}{UAS}{Unmanned Aerial System}
\newacronym{uav}{UAV}{Unmanned Aerial Vehicle}
\newacronym{udm}{UDM}{Unified Data Management}
\newacronym{udp}{UDP}{User Datagram Protocol}
\newacronym{udr}{UDR}{Unified Data Repository}
\newacronym{ue}{UE}{User Equipment}
\newacronym{uhd}{UHD}{\gls{usrp} Hardware Driver}
\newacronym{ul}{UL}{Uplink}
\newacronym{um}{UM}{Unacknowledged Mode}
\newacronym{uml}{UML}{Unified Modeling Language}
\newacronym{upa}{UPA}{Uniform Planar Array}
\newacronym{upf}{UPF}{User Plane Function}
\newacronym{urllc}{URLLC}{Ultra Reliable and Low Latency Communication}
\newacronym{usa}{U.S.}{United States}
\newacronym{usim}{USIM}{Universal Subscriber Identity Module}
\newacronym{usrp}{USRP}{Universal Software Radio Peripheral}
\newacronym{utc}{UTC}{Urban Traffic Control}
\newacronym{vim}{VIM}{Virtualization Infrastructure Manager}
\newacronym{vm}{VM}{Virtual Machine}
\newacronym{vnf}{VNF}{Virtual Network Function}
\newacronym{volte}{VoLTE}{Voice over \gls{lte}}
\newacronym{voltha}{VOLTHA}{Virtual OLT HArdware Abstraction}
\newacronym{vr}{VR}{Virtual Reality}
\newacronym{vran}{vRAN}{Virtualized \gls{ran}}
\newacronym{vss}{VSS}{Video Streaming Server}
\newacronym{wbf}{WBF}{Wired Bias Function}
\newacronym{wf}{WF}{Wired-first}
\newacronym{wlan}{WLAN}{Wireless Local Area Network}
\newacronym{osm}{OSM}{Open Source \gls{nfv} Management and Orchestration}
\newacronym{pnf}{PNF}{Physical Network Function}
\newacronym{drl}{DRL}{Deep Reinforcement Learning}
\newacronym{mtc}{MTC}{Machine-type Communications}
\newacronym{pt}{PT}{Plain Text}

\glsdisablehyper

\newcommand{\rev}[1]{\textcolor{black}{{#1}}}

\def\BibTeX{{\rm B\kern-.05em{\sc i\kern-.025em b}\kern-.08em
    T\kern-.1667em\lower.7ex\hbox{E}\kern-.125emX}}

\usepackage{verbatim}

\begin{document}

\title{Implementing and Evaluating Security in O-RAN: Interfaces, Intelligence, and Platforms \\
}


\author{
\IEEEauthorblockN{
Joshua Groen\IEEEauthorrefmark{1}, 
Salvatore D'Oro, 
Utku Demir, 
Leonardo Bonati, 
Michele Polese, \\
Tommaso Melodia,
Kaushik Chowdhury
} \\
\textit{Institute for the Wireless Internet of Things}, Northeastern University, Boston, MA, USA \\
\IEEEauthorblockN{
\IEEEauthorrefmark{1}groen.j@northeastern.edu,
\{f.last\}@northeastern.edu} \\ 
 }

\maketitle

\thispagestyle{plain}
\pagestyle{plain}

\begin{abstract}
The Open Radio Access Network (RAN) is a new networking paradigm that builds on top of cloud-based, multi-vendor, open and intelligent architectures to shape the next generation of cellular networks for 5G and beyond. While this new paradigm comes with many advantages in terms of observatibility and reconfigurability of the network, it inevitably expands the threat surface of cellular systems and can potentially expose its components and the \gls{ml} infrastructure to several cyber attacks, thus making securing O-RAN networks a necessity. In this paper, we explore security aspects of O-RAN systems by focusing on the specifications, architectures, and intelligence proposed by the O-RAN Alliance. 
We address the problem of securing O-RAN systems with \rev{a} holistic perspective, including considerations on the open interfaces used to interconnect the different O-RAN components, on the overall platform, and on the intelligence used to monitor and control the network.
%
For each focus area we identify threats, discuss relevant solutions to address these issues, and demonstrate experimentally how such solutions can effectively defend O-RAN systems against selected cyber attacks. 
This article is the first work in approaching the security aspect of O-RAN holistically and with experimental evidence obtained on a state-of-the-art programmable O-RAN platform, providing unique guideline for researchers in the field. 
\end{abstract}

\begin{picture}(0,0)(10,-450)
\put(0,0){
\put(0,0){\footnotesize This paper has been accepted for publication IEEE Network Magazine, DOI: 10.1109/MNET.2024.3434419.}
\put(0,-10){
\scriptsize \textcopyright~2024 IEEE. Personal use of this material is permitted. Permission from IEEE must be obtained for all other uses, in any current or future media, including}
\put(0, -17){
\scriptsize reprinting/republishing this material for advertising or promotional purposes, creating new collective works, for resale or redistribution to servers or lists,}
\put(0, -24){
\scriptsize or reuse of any copyrighted component of this work in other works.}
}
\end{picture}



\section{Introduction}

The increasing demand for multi-vendor, horizontally disaggregated systems in cellular \gls{ran} has exposed the limitations of traditional, closed, proprietary architectures. These systems are inadequate for meeting the stringent requirements of new services, prompting a shift towards open architectures that offer flexibility, programmability, and observability while reducing CAPEX and OPEX~\cite{polese2022understanding}. The Open \gls{ran} paradigm addresses these needs by promoting cloud-based, disaggregated, multi-vendor deployments with data-driven control, enabling flexible and customizable networks at lower costs.

\begin{figure}[t!]
\centering
  \includegraphics[width=\linewidth]{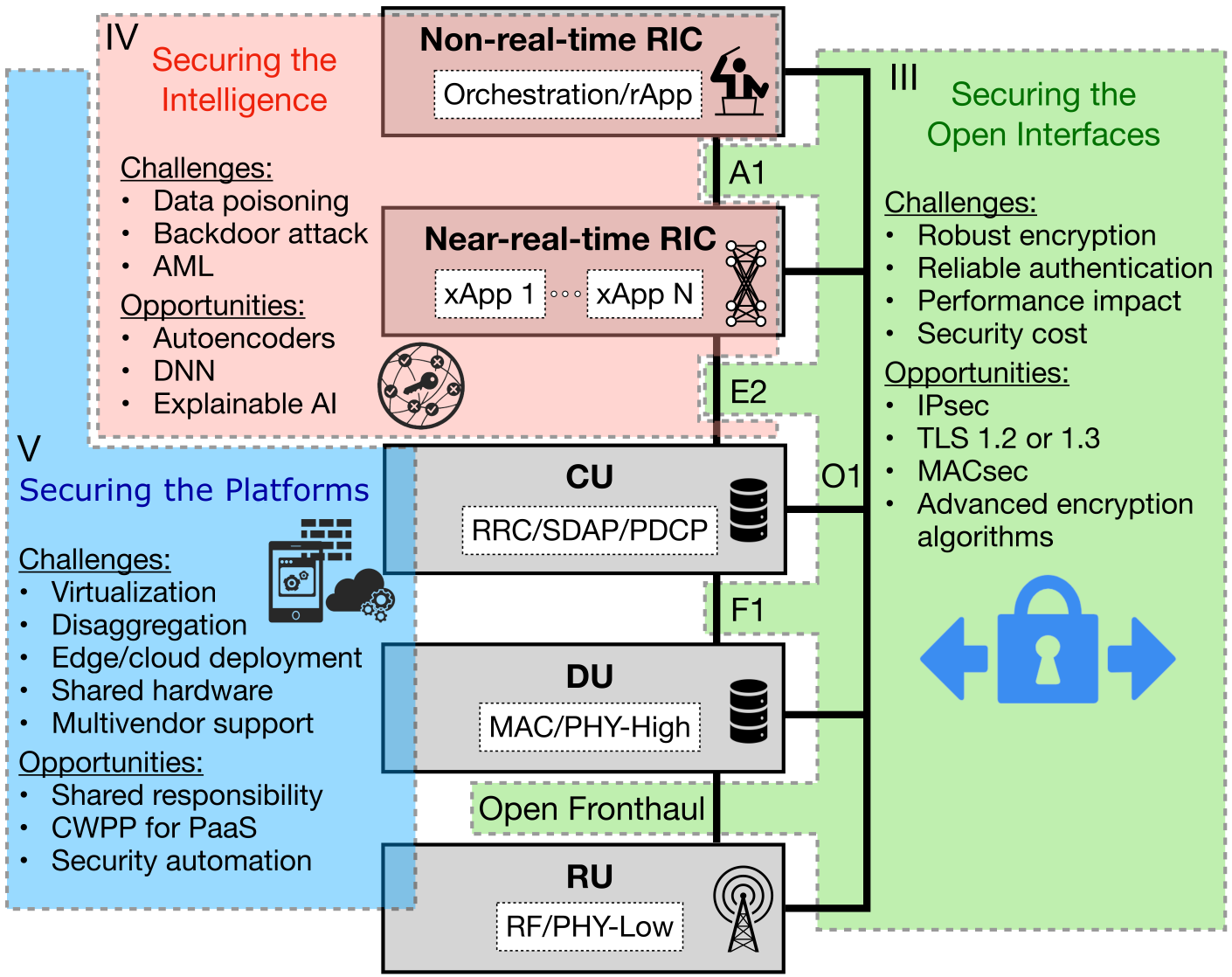}
  \caption {Challenges and opportunities to secure different groups of components in the O-RAN architecture, which also reflects the paper outline.}
  \label{fig:fig1}
\end{figure}

However, the \emph{openness} of O-RAN introduces security vulnerabilities, as fine-grained data extraction and network control capabilities can be exploited by attackers. Despite well-defined interface requirements, the impact of securing these interfaces is not fully understood. O-RAN's reliance on cloud-based virtualized functions and third-party \gls{ai}/\gls{ml} applications for closed-loop control further expands the threat surface. While there is ongoing research on securing \gls{ai}/\gls{ml} and cloud-based architectures, a comprehensive security solution for the entire O-RAN architecture remains an open challenge.

\begin{table*}
    \centering
    \resizebox{\linewidth}{!}{%
    \begin{tabular}{|c|l|c c c|c|} \hline
    Paper & Major Focus & Interfaces & Intelligence & Platforms & Implementation \\ \hline \hline
    \cite{polese2022understanding} & Comprehensive overview of O-RAN & \checkmark & \checkmark & \checkmark &  \\ \hline
    \cite{abdalla2022toward} & Survey of current capabilities and limitations of O-RAN & \checkmark & \checkmark & \checkmark &   \\ \hline
    \cite{mimran2022evaluating} & Proposed security evaluation ontology for O-RAN & \checkmark & \checkmark & \checkmark &   \\ \hline
    \cite{ramezanpour2022intelligent} & Proposed Zero Trust framework for 5G/6G~ & & \checkmark & \checkmark & \\ \hline
    \cite{ericcson} & Architectural security considerations for O-RAN & \checkmark &  & \checkmark &  \\ \hline
    \cite{dik2023open} & Open Fronthaul Security & \checkmark & & & \\ \hline
    \cite{groenCostOfficial} & Securing the E2 interface & \checkmark & & &  \checkmark \\ \hline
    \rowcolor[rgb]{1,1,0.694} This paper & Cost and effectiveness of implementing comprehensive security in O-RAN & \checkmark & \checkmark & \checkmark & \checkmark \\ \hline
    \end{tabular}
    }
    \caption{Survey of prior works showing the general focus of each paper, the security topics addressed, and if any solutions were implemented and evaluated in an O-RAN environment.}
    \label{tab: prior work}
\end{table*}

Security in O-RAN is still in its early stages. Tab.~\ref{tab: prior work} summarizes the focus, security topics addressed, and security implementations in the existing literature on O-RAN security. \cite{ramezanpour2022intelligent} explores zero-trust principles, adding network analytics and anomaly detection to authentication mechanisms. Abdalla {\em et al.}\cite{abdalla2022toward}, Mimran {\em et al.}\cite{mimran2022evaluating} and Polese {\em et al.}~\cite{polese2022understanding} identify potential threats and vulnerabilities but do not assess the cost of protection. To address these shortcomings, we identify three components within O-RAN—Open Interfaces, \gls{ai}-native closed loop control, and cloud-based platforms—that pose heightened risks and propose actionable strategies to mitigate these risks. Different from previous works,  we empirically evaluate the impact--in terms of cost and effectiveness--of implementing several of these measures. Specifically, the contributions of this article include: 
\begin{itemize}
    \item Reporting the observed effects on latency, throughput, and processing load from integrating encryption into the E2 interface.
    \item Assessing the efficacy of employing Autoencoders to mitigate potential attacks targeting the \gls{ai}.
    \item Identifying and detailing three pivotal principles of cloud security essential for safeguarding computing platforms in O-RAN deployments.
\end{itemize}

The organization of this paper, summarized in Fig.~\ref{fig:fig1}, is as follows. We first provide a brief background on the O-RAN architecture. Then, we discuss the open interfaces, followed by a baseline implementation evaluating the cost of securing them. Next, we demonstrate how to secure the intelligence and mitigate threats targeting \gls{ai}/\gls{ml} applications hosted on the near-RT \gls{ric}. After that, we provide an overview on securing platforms and disaggregated functions. Finally, we propose ways to advance security in O-RAN and draw our conclusions.




\section{O-RAN Primer}
\label{sec:ORANprimer}

The O-RAN Alliance defines a flexible and softwarized architecture (shown in Fig.~\ref{fig:fig1}) for the Open \gls{ran} ecosystem.
This architecture embraces the 3GPP 7.2x~functional split, where \gls{ran} functionalities are \emph{disaggregated} into \gls{cu}, \gls{du}, and \gls{ru}~\cite{ORANarch}.
These components carry out operations at different layers of the protocol stack. For example, the \gls{cu} performs RRC, PDCP, SDAP, and PDCP while the \gls{du} provides  RLC, MAC, and PHY-high functions. Both PHY-low and radio-frequency transmission are carried out at the \gls{ru}. 

Each of these components' functionalities can be controlled in software through exposed \glspl{api}.
Open and standardized interfaces connect these components both among them (e.g., the F1 interface connects \gls{cu} and \gls{du}, and Open Fronthaul interface connects \gls{du} and \gls{ru}), and to the \glspl{ric} (i.e., through the E2 and O1 interfaces).

These controllers are connected by the A1 interface and oversee the operations on the \gls{ran} nodes and enable closed control-loops that operate at different time scales through \gls{ai}/\gls{ml} applications.
The near-RT \gls{ric} acts on time scales between $10$\:ms and $1$\:s via applications called xApps, while the non-RT \gls{ric} on time scales above $1$\:s via rApps.
%
%
Finally, the applications running on the \glspl{ric} receive live \gls{ran} \glspl{kpi} from the \gls{ran} nodes through the E2 interface (in the case of xApps) and through the O1 interface (in the case of rApps). The \glspl{ric} can send control actions, e.g., to prioritize certain \glspl{ue} or to modify the scheduling policy of the \gls{du}, through the same control interfaces.
Interested readers are referred to~\cite{polese2022understanding} for a comprehensive overview of the O-RAN architecture and of its specifications.

\section{Securing the Open Interfaces}
\label{sec:SecOpenInterfaces}

\subsection{New Vulnerabilities}
The introduction of open interfaces that carry data between disaggregated components is a pillar of the O-RAN framework. However, the openness of the interface, combined with disaggregated nodes connected over open infrastructure, introduces expanded security vulnerabilities. In fact, some of the primary classes of threats arise from improper or missing ciphering of the data sent across these open interfaces and lack of proper authentication~\cite{groenCostOfficial, ramezanpour2022intelligent, mimran2022evaluating, abdalla2022toward}. For example, these vulnerabilities could be exploited with man-in-the-middle attacks to inject false KPI reports northbound to the near-RT RIC or inject malicious control actions southbound to the gNB.


The O-RAN Alliance recognizes these expanded threat vectors and each interface working group has published guidance to secure their respective interface. While the guidance varies greatly in the level of detail provided, in general each interface should provide confidentiality, integrity, and authentication. While this guidance is reasonable 
and there is general consensus that security is essential to the deployment of 5G O-RAN infrastructure~\cite{ericcson, OranWG3, DoD, groenCostOfficial, abdalla2022toward, ramezanpour2022intelligent,dik2023open}, there has been little study on the impact of securing the open interface in an O-RAN deployment. It is vital that an informed and risk-based approach is taken to adequately address security risks in O-RAN, while recognizing that any method for enhancing security, such as adding encryption, comes at a cost in terms of performance. This impact could reduce throughput or require additional equipment, affecting scalability and negatively impacting CAPEX. 

\subsection{Implementing IPsec on the E2 Interface}

To extract actionable insights, we thoroughly test and analyze the effects of adding encryption to the E2 interface as described in~\cite{groenCostOfficial}. We focus our efforts first on the E2 interface as this is the most mature interface, which makes it possible to add IPsec to this interface and observe the trade-offs between security robustness and performance.

While the E2 interface standards call for the use of IPsec, other open interfaces require the use of TLS 1.2, though use of TLS 1.3 is recommended. There are some differences in the way IPsec and TLS protocols initially establish a connection or \gls{sa}, however, we are confident we can extend the results generally without loss of accuracy because IPsec and TLS use the same underlying encryption and hashing algorithms once the \gls{sa} is established. TLS 1.3 has the fastest \gls{sa} establishment, making it hardest to confidently extrapolate the results directly from IPsec, but the general trends in performance still hold. Regardless of the security protocol used, system designers should only use modern secure encryption algorithms. 

After establishing a baseline performance without encryption, we add O-RAN-compliant encryption (as specified in~\cite{OranWG3}) to the E2 interface. We test multiple combinations of encryption algorithms and implementations, including AES128 (CBC, CCM), AES256 (CBC, CCM, GCM), and ChaCha20-Poly1305. 
IPsec, as configured in our test, provides confidentiality, integrity, replay protection, and authentication. With this configuration, IPsec adds at least $57$\:Bytes of overhead to each packet. However, because both AES and SHA2 require fixed input block sizes, padding may be added causing the overhead to further increase. For example, encrypting a 62 Byte selective acknowledgement (SACK) adds $76$\:Bytes for a total cypher text (CT) SACK length of $138$\:Bytes. 

\subsection{Theoretical Analysis}

To measure the impact to performance of adding security we examine delay and throughput as key performance indicators. In packet switched networks there are generally four sources of delay at each node along the path:
\emph{queuing} delay, \emph{propagation} delay, \emph{transmission} delay, and \emph{nodal processing} delay \cite{kurose1986computer}. 
\begin{itemize}
    \item \label{queuing}\emph{Queuing Delay:} In general, the queuing delay is not constant and depends on the packet arrival and departure rates. However, adding encryption to the link does not significantly impact the queuing delay. In other words, the difference in queuing delay with or without encryption is negligible. 
    \item \label{delay_prop} \emph{Propagation Delay:} The propagation delay is strictly a function of the physical length and propagation speed of the link and will remain constant regardless of encryption. 
    \item \emph{Transmission Delay:} The transmission delay is a function of the packet size (in bits), \(L\), and the link transmission rate, \(R\), which is defined as \(D_{trans}=L/R\). For any given system, \(R\) is fixed but \(L\) will increase to some extent with encryption. 
    \item \emph{Processing Delay:} Typically the processing delay is defined as the time required for intermediate nodes to examine the packet header and determine where to direct the packet \cite{kurose1986computer}. For this analysis, we include the encryption delay in the processing delay because it is essential to pass the payload to lower or higher OSI layers. 
\end{itemize}

While queuing and propagation delay are unaffected by encryption, transmission and processing delay are impacted. For this reason our analysis focuses on quantifying the increase in transmission and processing delays. While it is clear that there will be some impact, 
researchers should fully understand the expected trade-off for their system. 
Next, 
we outline a few experiments that system engineers can perform to better characterize the impact of security and ensure they plan for security by design.

\subsection{System Implementation Results}
\begin{figure}[htb]
    \centering
    \includegraphics[width=\linewidth]{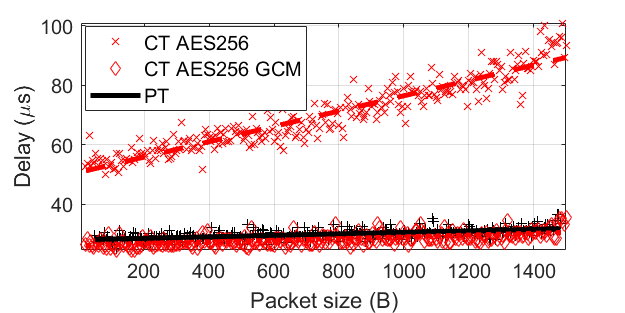}
    \caption{Processing delay as a function of packet size for PT and CT traffic.}
    \label{fig:process_delay}
\end{figure}

To quantify the effect of encryption for packets of various lengths, we use ping (ICMP echo) of various lengths to capture the network round trip time (RTT), which is twice the one-way delay. 
We subtract the fixed propagation delay and the known transmission delay and graph the processing delay in Fig. \ref{fig:process_delay}. We observe that the specific implementation of the AES algorithm greatly affects the additional processing delay. AES256-CCM causes the processing delay to increase with packet length while AES256-GCM has virtually no impact. 
\rev{Galois/Counter Mode (GCM) offers both confidentiality (via counter mode) and authentication (via arithmetic in the Galois field GF($2^{n}$)), where $n$ represents the key size. Since these operations can be executed in parallel, GCM delivers higher performance compared to modes like CBC, which necessitate sequential operation chaining~\cite{groen2024securing}.}
In either case, the processing delay difference between CT and \rev {\gls{pt}} is \(\Delta D_{proc} \leq 50 \mu s \) for all tested packet sizes. We can conclude that encryption has minimal impact on E2 traffic delay.

\begin{figure}[htb]
    \centering
    \includegraphics[width=\linewidth]{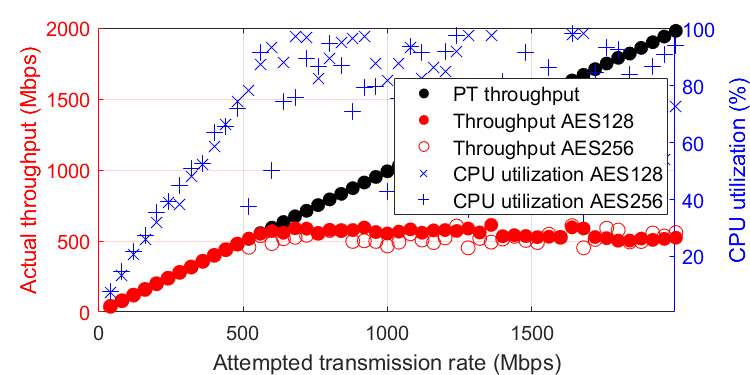}
    \caption{Measured throughput as a function of attempted transmission rate for PT and CT traffic. CPU utilization is the limiting factor for throughput with encryption.}
    \label{fig:throughput cpu}
\end{figure}

We also design an experiment to quantify the effect of encryption on the total traffic throughput. We use iperf3 to generate traffic at specific transmission rates for $10$\:s. 
In Fig. \ref{fig:throughput cpu} we see that the maximum encryption rate our system is capable of for AES256-CCM with SHA256 is a little under $600$\:Mbps. We also capture CPU utilization on the gNB for each attempted transmission rate in Fig. \ref{fig:throughput cpu} for both PT and CT, showing that encryption is very CPU intensive. The CPU utilization grows linearly until it reaches saturation. 

We test several other encryption algorithms and find a similar pattern for all. However, some algorithms are more CPU intensive than others. To illustrate the different impact a given encryption algorithm has on maximum throughput, we used iperf3 to send the maximum amount of traffic possible for $30$\:s. 
AES-CBC achieves 505 and 512 Mbps for key lengths 128 and 256 bits, respectively. AES-CCM achieves 573 Mbps for both key lengths of 128 and 256 bits. Cha-Cha20-Poly1305 achieves 989 Mbps, while AES-GCM achieves 1370 Mbps for all key lengths, 64, 128, and 256 bits.
%
The specific algorithm implementation is the most significant factor and AES-GCM vastly outperforms the other algorithm implementations. One key observation is that 
there is virtually no difference in performance based on key size (AES128 verse AES256). We encourage all system designers to use 256 bit key length as that provides higher security with virtually no impact to performance.

\subsection{Key to Securing Open Interfaces}

The key trade-off for securing O-RAN open interfaces is processing power. Any disaggregated gNB component must have enough CPU resources to manage all of its explicit functions. However, as the total traffic over the open interfaces increases,  CPU resources needed for encryption alone will also increase. One potential solution is to increase the total compute resources in the distributed gNB. For example, we compare OpenSSL benchmark tests on our Colosseum emulation environment built on Intel Xeon e5-2650 CPU to a desktop running an Intel Core i7-11700KF CPU and found that the desktop offers approximately a $3.5\times$ performance improvement. System designers could also add dedicated encryption hardware to the network stack to remove the burden from the CPU. 
A third option is to set strict limits on the amount of traffic that can be sent over the interfaces. 
\rev{This option requires careful consideration, as limiting interface throughput can significantly impact system performance. However, there may be safe reduction opportunities. For instance, in our system with 10 UEs, the E2 interface averages 3.5 Mbps with updates every 250 ms. However, this rate of updates may be higher than what is actually needed. By reducing the update frequency to every 333 ms the average throughput is reduced to 2.6 Mbps, decreasing link utilization by 25\%.}
In any case, system engineers must understand the amount of traffic expected across given interfaces and include the overhead of encryption in their system design. 







\section{Securing the Intelligence}
\label{sec:SecIntelligence}

O-RAN comes with the promise of \gls{ai}-native cellular networks where a fabric of \glspl{ric}, rApps and xApps monitor the network and take autonomous decisions to maximize performance and meet operator goals and intents~\cite{polese2022understanding}. By leveraging \gls{ai}/\gls{ml} techniques, rApps and xApps will enable applications ranging from control of network slicing and scheduling policies, traffic steering, mobility management, beamforming, and energy saving functionalities. 
However, these data-driven techniques are also known to have pitfalls and vulnerabilities that might expose rApps, xApps, \glspl{ric} and the entire network to a variety of security threats. 


\subsection{Vulnerabilities of AI/ML O-RAN Applications}

As rApps and xApps are the decision-making engines of O-RAN systems, it is fundamental that their decisions are unbiased, made with confidence, resilient against attacks or anomalies, and do not deviate from the goal of the network operator. Next, we highlight some vulnerabilities that affect \gls{ai}/\gls{ml} routines embedded in O-RAN applications.
\begin{itemize}
    \item \textit{Before Deployment:}~\gls{ai}/\gls{ml} solutions heavily rely upon the quality and quantity of data used to train them, which makes them vulnerable against \textit{data poisoning} and \textit{backdoor attacks}. Data poisoning aims at injecting misleading data into the data lakes to negatively impact the decisions made by data-driven xApps and rApps. One example is an adversarial \gls{ue} falsely reporting poor \gls{kpi} measurements (e.g., a low throughput level while experiencing high values). Any xApp or rApp trained using such biased and misleading data will result in sub-optimal decisions. Backdoor attacks, instead, aim at making \gls{ai}/\gls{ml} models sensitive against specific inputs, events, or \gls{kpi} patterns. For example, an attacker can develop an xApp controlling slicing policies with a backdoor such that the xApp makes decisions that are unfair toward certain subset of \glspl{ue} whenever a certain input triggers the backdoor, or takes actions that are in conflict with those of legitimate applications.

    \item \textit{After Deployment:}~Attacks might also affect real-time execution and inference of xApps and rApps. A well-known attack is \gls{aml}, where the attacker aims at either generating completely synthetic inputs, or at modifying an existing legitimate input via small and hard to detect perturbations. In O-RAN, these attacks can be extremely effective and can completely disrupt networking operations. This includes misleading a traffic classifier used to determine traffic type and corresponding \gls{qos}, bypassing anomaly detection mechanisms, steering control decisions away from their optimal, or leading the non-RT \gls{ric} into producing an inaccurate representation of network state. 
    Similarly, an attacker can potentially generate undetectable adversarial inputs that mislead an rApp into triggering wrong handover procedures, while at the same time causing an xApp to allocate fewer resources to the hand-off \glspl{ue}, resulting in unfairness and poor performance. 
    \end{itemize}


\subsection{Toward Reliable AI/ML O-RAN Applications}

Although the spectrum of attacks against data-driven xApps and rApps is broad, practitioners can rely upon a portfolio of well-established tools and solutions to protect the intelligence of the \glspl{ric}.

\begin{itemize}
    \item \textit{Preventing Attacks:}~The first step starts from designing and training xApps and rApps that are secure and robust against attacks by design. In this area, xApp/rApp developers can leverage several tools designed to minimize the impact of data variance on the output of \gls{ai}/\gls{ml} solutions embedded into O-RAN applications. A few examples are: projecting data into a latent space (e.g., \glspl{ae}), embedding synthetic adversarial and poisoned data into the training process (e.g., contrastive and adversarial learning, contaminated best arm identification, defensive distillation), and mitigating attack effectiveness via attention networks that aim at maintaining attention level on relevant features only and neglect adversarial influence.  

    \item \textit{Detecting Attacks:}~Although preventing attacks is always desirable,
    learning how to detect novel and increasingly sophisticated attacks becomes a necessity. Anomaly detection is the most relevant area of research that focuses on this issue and can help in designing secure O-RAN systems. 
    Among others, we mention techniques that detect anomalies by monitoring the distance between input data and the expected distribution (e.g., clustering- and distance-based algorithms, statistical methods), as well as deep neural networks that can identify unexpected data features that characterize anomalies. The former techniques could be used, for instance, to compare the decisions made by xApps/rApps at run-time with their statistical behavior, raising an anomaly warning in case these differ significantly. The latter, instead, could be leveraged to flag data with unexpected formats or patterns caused by anomalies in the network.

    \item \textit{Reacting to Attacks:}~In the case where an attack is successful and it has been detected, the best way to react is to understand what caused the attack, how it was performed and eventually learn to recover from it. Explainable \gls{ai} tools can be used to analyze the behavior of O-RAN applications, and to provide
    the non-RT \gls{ric} with information on which xApps and rApps were affected, what type of input produced the attack and why certain applications produced unexpected outputs. This information can then be used by the non-RT \gls{ric} to retrain model to eliminate vulnerabilities.
\end{itemize}

\subsection{Autoencoders for Threat Mitigation} \label{sec:ae_numerical}

In the following, we provide experimental results that illustrate how embedding \gls{ai}-based solutions like \acrfullpl{ae} into O-RAN applications can effectively help in mitigating adversarial attacks targeting the data-driven logic running therein.
We experimentally demonstrate that conventional autoencoders can deliver a good degree of protection against adversarial attacks, but more complex variations (e.g., variational, denoising and convolutional autoencoders) could be used for improved performance).

We consider the case of an xApp embedding a \gls{drl} agent trained to jointly control \gls{ran} slicing and scheduling policies of \gls{embb}, \gls{urllc} and \gls{mmtc} slices. \textcolor{black}{These agents are trained using a \gls{ppo} architecture with a reward that weights slice-specific \glspl{kpi}. Due to space limitations, we refrain from reporting details on the \gls{drl} design and training process, which can be found in~\cite{polese2022colo}. We consider two state configurations: } direct feeding of \glspl{kpi} to the \gls{drl} agent, and preprocessing \glspl{kpi} through an \gls{ae} for dimensionality reduction and outlier suppression before feeding them to the \gls{drl} agent. We consider the case of \textcolor{black}{black-box adversarial attacks. The attacker uses an attack vector consisting of corrupted \gls{kpi} measurements generated by taking legitimate \gls{kpi} data and perturbing them with additive random noise} following a normal distribution with mean equal to the \gls{kpi} value being perturbed and variance $\sigma$.

\begin{figure}[htb]
    \centering
    \includegraphics[width=\linewidth]{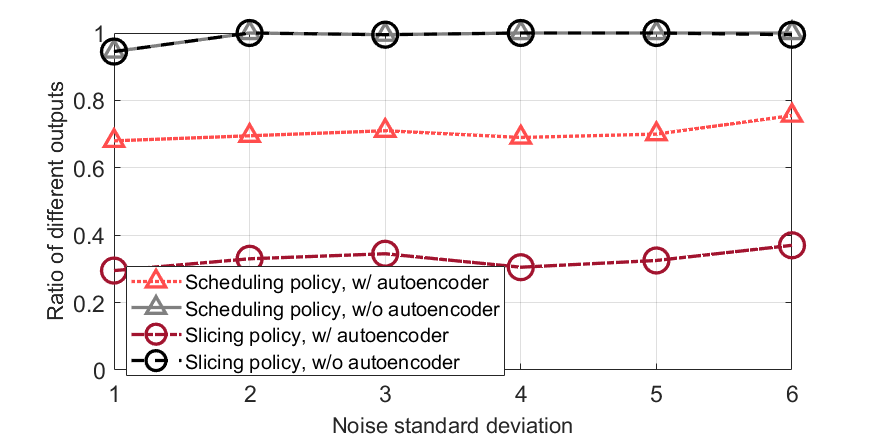}
    \caption{Percentage of actions deviating from the expected outcome after the attack with and without the autoencoder.}
    \label{fig:difference_actions}
\end{figure}

To evaluate the effectiveness of the attack, we vary the standard deviation of such a random variable and run 200 independent runs. For each run, the \gls{drl} agent is fed first with \glspl{kpi} reported by a cellular base station, and then by those affected by the attack.
In Fig.~\ref{fig:difference_actions}, we show the percentage of actions that deviate from the intended original actions taken by the \gls{drl} agent after attack. We perform this analysis by considering two cases. In the first case, we feed the perturbed \glspl{kpi} directly to the \gls{drl} agent, while in the second case the \glspl{kpi} are first processed by an \gls{ae}. Our results show that the \gls{ae} always results in fewer deviations from the intended actions even in the case of high noise variance. On the contrary, the case where no \gls{ae} is used results in more than 95\% deviation from the intended action. Moreover, we notice that the most affected action is the scheduling policy, which deviates 70\% of times, while the slicing policy is affected only 30\% of times.


\begin{figure}[htb]
    \centering
    \includegraphics[width=\linewidth]{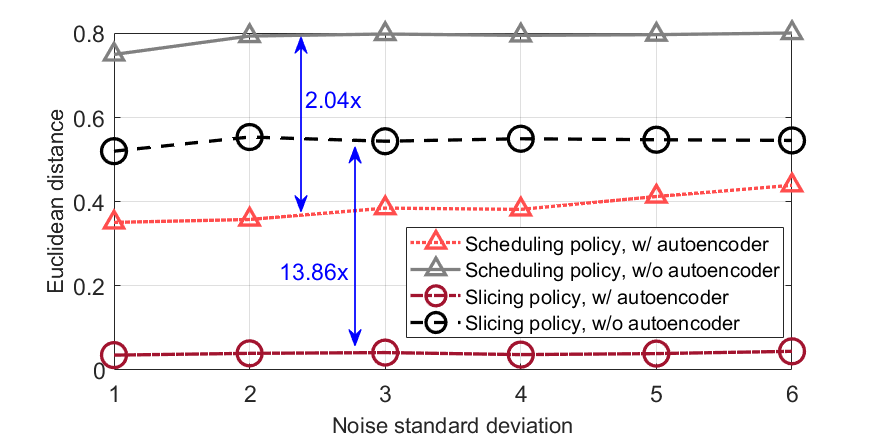}
    \caption{Action distance comparison after attack with and without autoencoder. Scheduling policies are enumerated as $\{0,1,2\}$, while slicing actions consist of three integer numbers (one per slice) whose sum must be equal to 50 (i.e., the number of Physical Resource Blocks (PRBs) to assign to each slice).}
    \label{fig:distance_actions}
\end{figure}

In Fig.~\ref{fig:distance_actions}, we investigate how much an attack can steer the xApp/rApp agent away from the intended behavior, i.e., how distant from the intended outcome the actions taken by the \gls{drl} agent are, by reporting the Euclidean distance between the intended action and the one taken by the \gls{drl} agent.
Scheduling policies are enumerated as $\{0,1,2\}$, while slicing actions consist of three integer numbers (one per slice) whose sum must be equal to 50 (i.e., the number of Physical Resource Blocks (PRBs) to assign to each slice). We measure the Euclidean distance between the intended action and the one taken by the \gls{drl} agent, and in Fig.~\ref{fig:distance_actions} we report the normalized Euclidean distance.
Results show that the \gls{ae} reduces the distance (and therefore the effectiveness of the attack) between scheduling decisions made after the attack by approximately $2\times$, and by more than $13\times$ with respect to the slicing policies.

\section{Securing the Platforms}
\label{sec:SecPlatformSystems}

One of the defining characteristic of O-RAN is the shift toward virtualization of network functions \cite{mimran2022evaluating, polese2022understanding, ericcson, abdalla2022toward}. This shift supports several goals including multi-vendor support, disaggregation of nodes, scalability, and open source development. In practice, this is accomplished in a way that follows cloud service deployments. Specifically, network functions are hosted and executed in virtual machines (or containers) managed by a hypervisor or host operating system (OS) to provide resource management, performance optimization, and access to common interfaces, among other services. While virtualization and cloud-based deployments result in a much more agile and flexible network infrastructure, they inevitably result in a much larger threat surface area.

\subsection{Cloud Security Risks}

One of the key components of any cloud infrastructure is common hardware components shared by multiple software instances. In the context of cellular communications, this introduces a new risk as isolation between applications is logical only, without physical isolation across hardware \cite{ericcson}. 
Additionally, the increased number of layers controlled by different parties in a virtual system provides additional threat vectors. An application must rely on multiple lower layers to perform specific security functions. For example, the host operating system has access to all RAM memory, disk volumes mounted on virtual machines, and containers \cite{ericcson}. As another example the hypervisor must access hardware security functions and pass those functions to the application. This means that the application must be able to trust both the hardware itself and the virtualization layer. To fully trust an application, one needs to trust all the layers in the stack. 

There are several existing cloud security architectures built on the concept of shared responsibility \cite{lane2017managing}. In general, the idea of shared responsibility means the cloud service provider is responsible for the security of all the components necessary to operate the cloud service, while the customer is responsible for protecting their data. In cloud security, the specific breakdown in responsibility is often a function of the type of cloud service (i.e., Infrastructure as a Service versus Platform as a Service (PaaS)). An O-RAN deployment most closely resembles a PaaS deployment. One of the security services typically used for a PaaS deployment is a Cloud Workload Protection Platform (CWPP). The objective of the CWPP is to keep the application, whether a VM or container, secure. System engineers should review PaaS security services for valuable lessons when designing O-RAN security and applications.

\subsection{O-RAN Specific Security Recommendations}

Unlike typical cloud deployments, one of the features of O-RAN is enabling multiple vendors to provide different functions. This creates additional challenges to principles such as supply chain security, secure service administration, and personnel security. This also means traditional appliances such as external security gateways are infeasible in terms of both CAPEX and OPEX to deploy between each and every component of a distributed O-RAN system. Some of these challenges could be \rev{mitigated} through a strong governance framework and appropriate visibility and auditing. On the other hand, O-RAN's stated goal of increasing automation and using AI and ML naturally supports principles such as security automation and flexibility. 

Regardless of the cloud service architecture, there are many existing cloud security frameworks that can be applied. For example, the UK National Cyber Security Centre published 14 Cloud Security Principles~\cite{UKNatCyberSec}, while Amazon AWS uses 7 Design Principles. A full discussion of all of these principles is beyond the scope of this paper, but we will address a few of the key principles that should be applied to O-RAN systems. 

$\bullet$~\textbf{Separation Between Customers.} O-RAN deployments must ensure separation between customers so that a malicious or compromised component cannot affect the service or data provided by another component. This responsibility belongs to both cloud service providers to secure the hypervisor or host OS and on developers providing new functionality, such as ensuring xApps do not leak customer information to other entities. 

$\bullet$~\textbf{Visibility and Auditing.} O-RAN deployments should also embrace visibility and auditing so that it is clear to all parties who is responsible for what function and the actions taken by each component can be traced. While visibility is inherently present to a degree in open source deployments, auditing is often overlooked. Further, \rev{as more complicated intelligence is introduced to closed loop control}, understanding which component directed a particular action becomes both more difficult and more necessary to secure each component. One way to achieve this is to ensure the \glspl{ric} maintains logs of the control messages sent by each xApp and rApp.  

$\bullet$~\textbf{Security by Design.} O-RAN must embrace security by design at all layers of the O-RAN stack. Each component in Fig.~\ref{fig:fig1} should be built with security in mind from the beginning. This can be especially challenging at the lower levels such as the RU and DU where performance is at a premium. There is always a trade off between security and performance. However, starting with security as one of the necessary objectives for each block allows researchers to optimize the solutions before deployment and before an attack.

\section{Advancing the Security Framework}


Our study serves as a foundational reference for future O-RAN security research, highlighting vulnerabilities in its components and demonstrating the efficacy of proposed methods through practical implementations. However, achieving a completely secure O-RAN deployment requires further research to build integral, in-depth, comprehensive security. Future efforts should focus on securing the Open Fronthaul and implementing a Zero Trust Architecture.

$\bullet$~\textbf{Open Fronthaul.} The Open Fronthaul includes the User plane, Control plane, Sycnhronization plane, and Management plane. While each of these planes has unique security requirements, all of them require low latency. For example, the synchronization plane does not require confidentiality, but authentication and integrity are crucial for timing packets crossing an insecure switched network. A man-in-the-middle attack could insert, delete, modify, or replay these messages causing a significant degradation or outage of service. Ensuring each plane provides the appropriate security function without impacting the quality of services is an open problem. 

$\bullet$~\textbf{Zero Trust Architecture.} Existing security models often assume a high level of trust among entities, which doesn't align well with the O-RAN framework~\cite{ramezanpour2022intelligent}. In contrast, Zero-Trust Architectures (ZTA) align well with emerging 5G network infrastructure~\cite{DoD}. ZTAs offer security assurances by adopting a data-centric model rather than traditional perimeter-based methods. This paradigm shift redefines networks as platforms for distributed data management, emphasizing data protection at every stage of the data cycle.

O-RAN presents an opportunity for implementing ZTA. Key ZTA elements include dynamic risk assessment and continuous trust evaluation. Researchers can leverage the \gls{ai}/\gls{ml} integration inherent in O-RAN for these tasks. Further development and implementation of ZTA xApps and rApps should focus on real-time monitoring, risk assessment, and ensure the availability, integrity, and confidentiality of a dynamic, multi-vendor, virtual O-RAN platform.

\section{Conclusions}
\label{sec:conclusions}

In this article, we provide a holistic guideline for securing an O-RAN system by classifying the potential targets as open interfaces, data-driven decision making components, i.e. intelligence, and cloud-based platforms. We indicate possible threats and suggest how these can be addressed using various methods. We then provide baseline implementation results for open interfaces and intelligence in order to showcase the impact of the few suggested solutions. 

Specifically, we observed that the IPsec protocol, which is the recommended security protocol for E2 interface by O-RAN Alliance, has minimal affect on delay for encrypting packets in the open interfaces, as the encryption adds a delay of less than $50\mu s$. However, encryption comes with a computation cost, requiring higher CPU for increasing throughput. In our environment, CPU utilization is the key cost to encryption.

Threats to the intelligence (i.e., \gls{ai}/\gls{ml} enabled \glspl{ric}) can come before and after deploying ML models. Security measures for the intelligence include preventing, detecting, and reacting to attacks. In this paper, we showcased an implementation to mitigate data poisoning attacks against an xApp that runs a DRL model using an autoencoder layer before the input to the DRL. Our experiments revealed that autoencoders can decrease the unexpected ML outputs for scheduling and slicing by 30\% and 70\%, respectively.

Lastly, we provide security measure directions on the overall platform, where we suggest researchers to focus on shared cloud responsibility. In this model, the service providers and users are responsible for securing the service and data, respectively, which can be realized through a CWPP, for example. We also highlight three key principles for O-RAN specific cloud deployment, including separation between customers, visibility and auditing, and security by design.


\section*{Acknowledgment}
This article is based upon work partially supported by Qualcomm Inc. and by the U.S.\ National Science Foundation under grants CNS-1925601, CNS-2112471, CNS-1923789 and CNS-2120447.
   
\balance
\bibliographystyle{IEEEtran}
\bibliography{main}

\begin{IEEEbiographynophoto}
{Joshua Groen} is a Ph.D. student at Northeastern University. Previously he worked in the US Army Regional Cyber Center – Korea as the Senior Network Engineer. He received his MS ('17) and BSE ('07) in Electrical Engineering respectively from the University of Wisconsin and Arizona State University. His research interests include wireless communications, security, and machine learning.
\end{IEEEbiographynophoto}

\vspace{-0.5cm}

\begin{IEEEbiographynophoto}
{Salvatore D’Oro} is a Research Assistant Professor at Northeastern University. He received his Ph.D. from the University of Catania in 2015. He serves on the Technical Program Committee of IEEE INFOCOM. His research focuses on optimization and learning for NextG systems.
\end{IEEEbiographynophoto}

\begin{IEEEbiographynophoto}
{Utku Demir} is a Postdoctoral Research Fellow at Northeastern University, where he is supported by the Roux Institute's Experiential AI program. He received his PhD from the University of Rochester in 2020. His research interests lie in the areas of wireless communications, mobile networks, signal processing, and machine learning.
\end{IEEEbiographynophoto}

\begin{IEEEbiographynophoto}
{Leonardo Bonati} is an Associate Research Scientist at Northeastern University. He received his Ph.D. from Northeastern University in 2022. His research focuses on softwarized NextG systems.
\end{IEEEbiographynophoto}

\begin{IEEEbiographynophoto}
{Michele Polese} is a Research Assistant Professor at Northeastern University. He obtained his Ph.D. from the University of Padova in 2020. His research focuses on architectures for wireless networks.
\end{IEEEbiographynophoto}

\begin{IEEEbiographynophoto}
{Tommaso Melodia} is the William Lincoln Smith Professor at Northeastern University, the director of the Institute for the Wireless Internet of Things, and the director of research for the PAWR Project Office. He received his PhD from the Georgia Institute of Technology in 2007. His research focuses on wireless networked systems.
\end{IEEEbiographynophoto}

\begin{IEEEbiographynophoto}
{Kaushik Chowdhury} is a Professor at Northeastern University, Boston, MA. He received his PhD from Georgia Institute of Technology in 2009. His current research interests involve systems aspects of machine learning for agile spectrum sensing/access, unmanned autonomous systems, programmable and open cellular networks, and large-scale experimental deployment of emerging wireless technologies.
\end{IEEEbiographynophoto}



\end{document}